# Strain engineering of Schottky barriers in single- and few-layer MoS$_2$ vertical devices.

*Jorge Quereda[1]\*, Juan José Palacios[1,3], Nicolás Agräit[1,2], Andres Castellanos-Gomez[2] and Gabino Rubio-Bollinger[1,3,4]*.

[1]Departamento de Física de la Materia Condensada. Universidad Autónoma de Madrid, Madrid, E-28049, Spain. [2]Instituto Madrileño de Estudios Avanzados en Nanociencia (IMDEA – Nanociencia), E-28049, Madrid, Spain. [3]Instituto de Ciencia de Materiales Nicolás Cabrera, E-28049, Madrid, Spain. [4]Condensed Matter Physics Center (IFIMAC), Universidad Autónoma de Madrid, E-28049, Madrid, Spain.

\* e-mail: jorge.quereda@uam.es

**Abstract:** We study the effect of local strain in the electronic transport properties of vertical metal-atomically thin MoS$_2$-metal structures. We use a conductive atomic force microscope tip to apply different load forces to monolayer and few-layer MoS$_2$ crystals deposited onto a conductive indium tin oxide (ITO) substrate while measuring simultaneously the *I-V* characteristics of the vertical tip/MoS$_2$/ITO structures. The structures show rectifying *I-V* characteristics, with rectification ratios strongly dependent on the applied load. To understand these results, we compare the experimental *I-V*s with a double Schottky barrier model, which is in good agreement with our experimental results and allows us to extract quantitative information about the electronic properties of the tip/MoS$_2$/ITO structures and their dependence on the applied load. Finally, we test the stability of the studied structures using them as mechanically tunable current rectifiers.

**Keywords:** two-dimensional materials, molybdenum disulfide (MoS$_2$), strain engineering, metal-semiconductor junction, Schottky barrier.

## 1. Introduction

Modifying the electronic properties of a material through mechanical strain is a powerful strategy to improve the performance of electronic devices [1-6]. This strategy has been recently applied with high success to atomically thin MoS$_2$, a two-dimensional semiconductor with huge technological potential. The attractive electronic and optoelectronic properties of MoS$_2$ have been used in the design of several electronic and optoelectronic devices [6-9], and more recently, also in the development of atomically thin memristors [10]. Due to their high Young's modulus [11] and elastic strain limit [12], atomically thin MoS$_2$ crystals can withstand large strains before rupture or inelastic relaxation take place [13, 14]. Interestingly, recent literature stated that the electronic bandgap of atomically thin MoS$_2$ can be reduced to a large extent by applying local strain to its crystal lattice [15-17], which opens the possibility of modulating the electrical conductivity of MoS$_2$ by local mechanical deformation.

In this work, we explore experimentally the effect of strain in the electron transport through vertical metal/atomically thin MoS$_2$/metal junctions, using a conductive AFM tip to apply different load forces to monolayer and few-layer MoS$_2$ crystals deposited onto a conductive indium tin oxide (ITO) substrate while measuring simultaneously the *I-V* characteristics of the vertical tip-MoS$_2$-ITO structures. Remarkably, even when the MoS$_2$ crystal is just one layer thick, the structure shows a strong rectifying *I-V* characteristic, indicating that two metal-semiconductor Schottky barriers [18] are formed at the tip/MoS$_2$ and MoS$_2$/substrate interfaces. Furthermore, we prove how the formed Schottky barriers and the electrical conductivity of the MoS$_2$ can be tailored applying local mechanical stress to the crystal with the AFM tip.

## 2. Results and discussion

Figure 1a shows an optical image of an atomically thin MoS$_2$ crystal on the ITO substrate. The MoS$_2$ flakes were exfoliated from a commercial n-doped MoS$_2$ crystal (SPI Supplies, 429ML-AB) by micromechanical cleavage with viscoelastic stamps and later transferred onto the ITO substrate, as explained in previous publications [19, 20], as well as in the Supplementary Information of this work. ITO was chosen as substrate due to its low sheet resistance and high transparency [21-23], useful for many applications, such as light-emitting diodes, solar cells or touch screens.

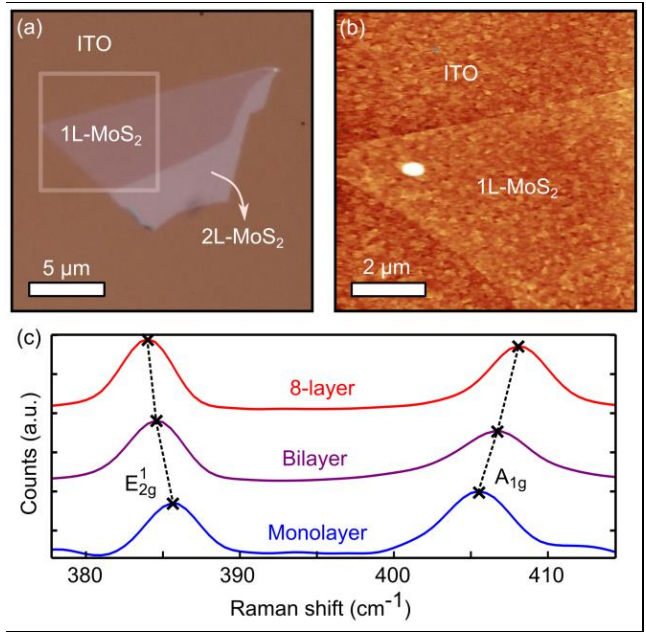

**Figure 1.** (a) Optical image of a MoS$_2$ flake with single layer and bilayer regions, deposited onto an indium tin oxide substrate (b) AFM topography image of the region marked by a square in (a). (c) Raman spectra showing the thickness-dependent frequency shift ($\Delta f$) between the E$^1_{2g}$ and A$_{1g}$ peaks in a monolayer ($\Delta f$ =19 cm$^{-1}$), a bilayer ($\Delta f$ =22 cm$^{-1}$) and an 8-layer ($\Delta f$ =23.5 cm$^{-1}$) thick flake.

Figure 1b shows a contact-mode AFM image of the same crystal shown in Figure 1a, with a monolayer region. We use Raman spectroscopy as a complementary method to determine the thickness of the MoS$_2$ flakes [24, 25]. Figure 1c shows the A$_{1g}$ and E$^1_{2g}$ peaks of the MoS$_2$ Raman Spectra for a monolayer, a bilayer and an 8 layers thick flake. As reported by Lee *et al*. [24], the frequency difference $\Delta f$, between these two peaks depends monotonically on the number of MoS$_2$ layers. We obtain $\Delta f = 19$ cm$^{-1}$ for the single layer, 22 cm$^{-1}$ for the bilayer and 24 cm$^{-1}$ for the 8 layers thick flake, consistent with the values found in literature.

Next, we investigate the *I-V* characteristics of the ITO-MoS$_2$-tip structures as well as their dependence on the tip-flake load force. The semiconductor MoS$_2$ flake is sandwiched between the ITO substrate and a conductive diamond AFM tip (ETALON series HA_HR_DCP, from NT-MDT), as depicted in Figure 2a, and two metal-semiconductor junctions are formed at the ITO-MoS$_2$ and the MoS$_2$-tip interfaces. A conductive diamond tip was chosen because of its high wear resistance, which allowed to identify the MoS$_2$ crystals by tapping mode AFM and then measure *I-V*s using the same tip. We carried out the same measurements using a platinum-coated AFM tip and obtained similar results, but the metallic coating of the tip was easily damaged, making the experiment hard to reproduce. Figure 2b shows the band diagram of the structure, where a Schottky barrier of height $\varphi_{tf}$ and ideality factor $n_{tf}$ is formed at the tip-flake junction and another Schottky barrier of height $\varphi_{fs}$ and ideality factor $n_{fs}$ is formed at the flake-substrate junction.

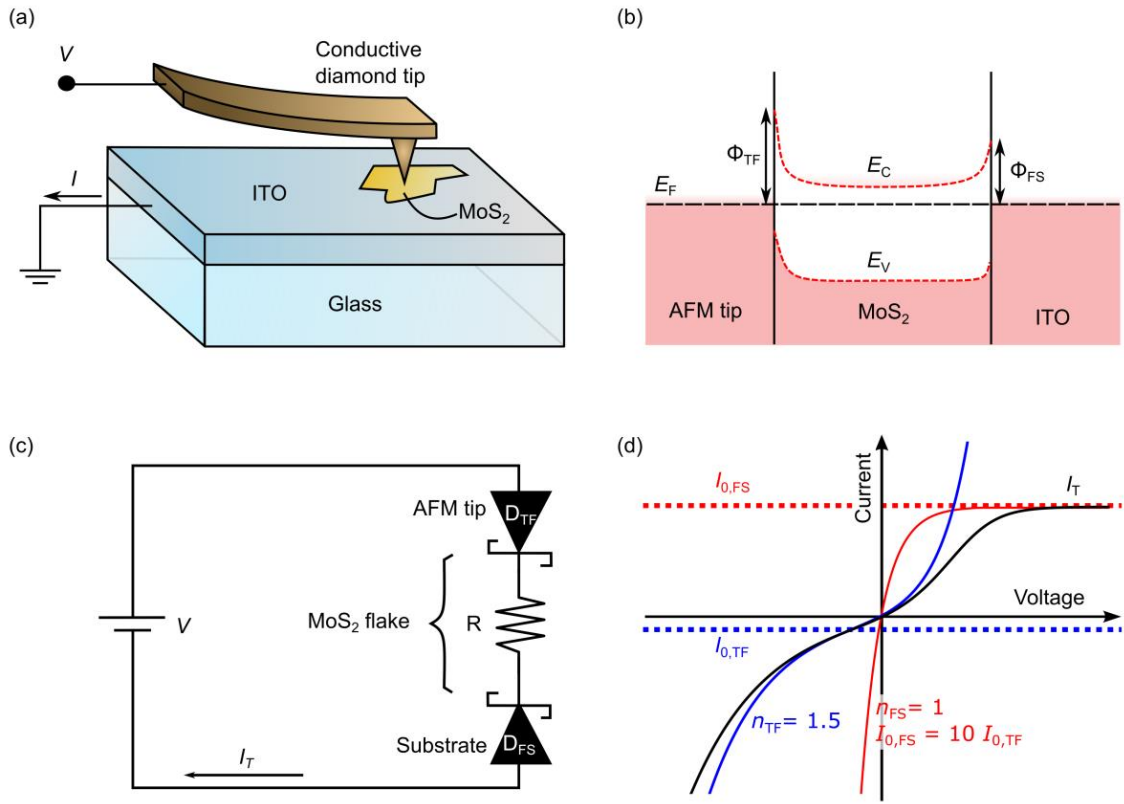

**Figure 2.** (a) Schematic of the experimental setup: We apply a voltage between the tip and the ITO substrate to measure the *I-V* characteristic of the MoS$_2$ crystal in the out-of-plane direction under different tip-flake contact forces. (b) Band diagram of the tip/MoS$_2$/ITO system showing the Schottky barriers, $\Phi_{TF}$ and $\Phi_{FS}$, formed at the tip-MoS$_2$ and MoS$_2$-ITO interfaces. (c) Circuit model. The tip-flake and flake-substrate interfaces are represented as two Schottky diodes, D$_{TF}$ and D$_{FS}$ respectively, connected in series in a back-to-back configuration. A series resistance, R, connected between the two diodes accounts for the finite conductivity of the MoS$_2$ crystal. (d) Typical *I-V* characteristic of the circuit (black curve) and of the individual tip-MoS$_2$ (blue) and MoS$_2$-substrate (red) barriers, considering different ideality factors, $n_{TF} > n_{FS} = 1$ and saturation currents, $I_{0,FS} = 10 I_{0,TF}$, for the two barriers.

Figure 2c shows a circuit model for the system: The tip-flake and flake-substrate interfaces are represented as Schottky diodes connected in a back-to-back geometry [26] (as in each interface the semiconductor side of the junction is facing an opposite direction). A resistance, *R*, connected in series between the two diodes accounts for the finite conductivity of the MoS$_2$ crystal.

The back-to-back Schottky diode picture and its associated band diagram in Fig. 2b must be interpreted with some care. Since the flakes studied are formed by few layers (some even by single one as shown below), there is no room for a standard band-bending picture in the perpendicular direction, i.e., only as a function of the z direction. Perpendicular injection does not necessarily imply that the current only flows perpendicularly. In fact, since the intralayer sheet resistivity of the individual layers is expected to be about 200 times smaller than the interlayer resistivity [27, 28], current, as injected from the tip for instance, is expected to flow in-plane and spread before reaching the bottom electrode. In other words, the axis along which the potential profile has been schematically depicted in Figure 2b must not be exclusively understood as the perpendicular direction to the layers. This way, even for a single layer, the implicit back-to-back Schottky diode picture still applies since the Schottky barriers are not on top of each other. To support this, and following the transmission line model commonly used to understand current injection into 2D materials [29, 30], we have modeled our system with a classical resistor network (see details in the Supporting Information). There it is shown how, as expected, the current and the voltage drop progressively spreads in plane from top to bottom across the layers.

According to the thermionic emission theory, the current $I_i$ through a metal-semiconductor junction (including image force effects) [31] is given by

$$I_i(V_i) = I_{0,i} \exp\left(\frac{qV_i}{n_i k_B T}\right)\left[1 - \exp\left(-\frac{qV_i}{k_B T}\right)\right] ; \quad (1)$$

$$I_{0,i} = A_c A^* T^2 \exp\left(\frac{q\varphi_i}{k_B T}\right), \quad (2)$$

where $I_{0,i}$ is the saturation current of the junction, $q$ is the electron charge, $V_i$ is the applied bias voltage, $n_i$ is the ideality factor of the diode, $k_B$ is the Boltzmann constant, $T$ is the temperature, $A_c$ is the metal-semiconductor contact area, $A^*$ is the effective Richardson constant and $\varphi_i$ is the energy of the Schottky barrier. It is common in literature to use the Schottky-Mott rule to estimate the value of $\varphi_i$ as the difference between the work function of the metal and the electron affinity of the semiconductor [31]. However, care needs to be taken in this case when using the Schottky-Mott rule, since MoS2 crystals have been reported to be intrinsically doped by chemical impurities and Sulphur vacancies [32, 33]. In addition to a Fermi level change, these can certainly modify the Fermi level at the metal-semiconductor interface [34], which is known to lead to strong discrepancies between the experimental energy of the Schottky barrier and the value predicted by the Schottky-Mott rule. Further, it remains unclear how the fermi level pinning could be affected by the applied strain.

For the complete circuit we have:

$$I_{\text{tf}}(V_{\text{tf}}) = -I_{\text{fs}}(V_{\text{FS}}) \equiv I_{\text{T}}(V) \; ; \tag{3}$$

$$V = V_{\text{tf}} + V_{\text{fs}} + I_{\text{T}}(V) \cdot R \, , \tag{4}$$

where $I_{\text{T}}(V)$ is the total current along the circuit and the subscripts *tf* and *fs* indicate the tip/flake interface and the flake/substrate interface respectively. Replacing $V_{\text{fs}}$ in Equation 3 by its expression derived from Equation 4 and re-ordering terms we get the equation:

$$I_{\text{tf}}(V_{\text{tf}}) = I_{\text{fs}}(V - V_{\text{tf}} - I_{\text{tf}}(V_{\text{tf}}) \cdot R) \, , \tag{5}$$

that must be solved by numerical methods, as detailed in the Supporting Information, to obtain $V_{\text{tf}}$ for each given applied voltage, $V$. Then, we can get the total current $I_{\text{T}}(V) = I_{\text{tf}}(V_{\text{tf}})$ from Equation 1 for any given values of the saturation currents, $I_{0,\text{tf}}$ and $I_{0,\text{fs}}$, the ideality factors, $n_{\text{tf}}$ and $n_{\text{fs}}$, and the series resistance $R$.

Figure 2d shows a calculated *I-V* characteristic $I_{\text{T}}(V)$ for the case $I_{\text{fs}} = 10 \, I_{\text{tf}}, n_{\text{tf}} = 1.5, n_{\text{fs}} = 1$ and $R = 0$. The blue and red curves in the figure are the *I-V*s of the tip-flake and flake-substrate Schottky barriers, $I_{\text{tf}}(V)$ and $I_{\text{fs}}(V)$, respectively. In this case, at positive voltages, the barrier formed at the flake-substrate interface is in saturation regime, limiting the total current of the circuit to a maximum value of $I_{0,\text{fs}}$. At negative voltages, however, the total current is limited by the tip-flake barrier and increases monotonically, even after reaching the saturation current, $I_{0,\text{tf}}$, due to the higher ideality factor of this barrier, $n_{\text{tf}}$. Note that for low positive voltages, the *I-V* of the circuit shows a distinctive curvature change, not present in the single-diode *I-V* characteristics, which is caused by the interplay between the two barriers.

To measure the *I-V* characteristics we contact the MoS₂ crystals transferred onto the ITO substrate with the conductive AFM tip, as shown schematically in Figure 2a, and then apply a bipolar voltage ramp of ±1V between the tip and the substrate. The applied tip-flake load force was calculated from the vertical displacement of the cantilever using its nominal elastic constant ($k = 16$ N/m). Figure 3a shows the measured *I-V*s in an 8 layers thick MoS₂ flake under different tip-flake load forces (colored markers), as well as their least-square fits to the theoretical model (black lines), performed using the series resistance, $R$, the ideality factors, $n_i$, and the saturation currents, $I_{0,i}$ as fit parameters. An 8 layers thick flake was selected because it is thick enough to avoid the flow of tunneling current between the tip and the substrate. The resulting *I-V* characteristics show a strong rectifying behavior, as well as a remarkable sensitivity to the applied force, with a huge increase of the rectification ratio $|I(-1V)/I(1V)|$, from ~1 at 0 nN to 18 at 80 nN.

We further study the effect of the tip-flake load force in the MoS₂ resistance, as well as in the ideality factors and barrier heights of the Schottky barriers, extracted from the fitting of the *I-V* characteristics to the model. Figure 3b shows the force-dependent MoS$_2$ resistance under load forces ranging from 0 to 80 nN. As the tip load is increased, *R* decreases from 8.3 MΩ to 0.6 MΩ, which we attribute to an increase of the interlayer coupling due to a strain-induced reduction of the interlayer distance. A similar trend has been theoretically described for MoS$_2$ crystals under vertical strain [35], as well as under biaxial horizontal strain [36].

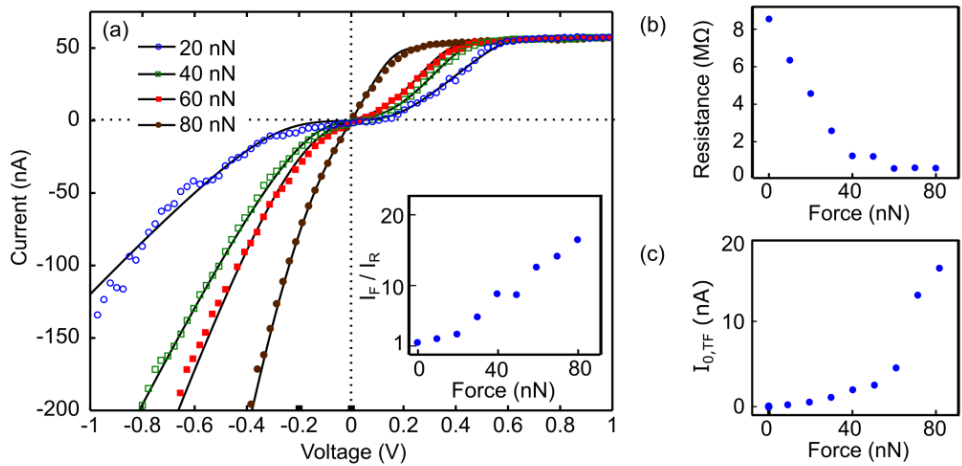

**Figure 3.** (a) Measured *I-V* characteristics of an 8 layers thick MoS$_2$ flake under four different tip-flake contact forces: 20, 40, 60 and 80 nN. The black lines are least square fittings to the model. Inset: Force-dependent rectification ratio measured at ± 1V. (b) Force-dependent resistance, *R*, of the MoS$_2$ crystal, extracted from the fitting of the *I-V* curves. As the contact force is increased, *R* decreases from 8.3 MΩ to 0.6 MΩ, indicating a force-induced reduction of the interlayer distance and, therefore, a higher interlayer coupling. (c) The saturation current of the tip-flake Schottky barrier, $I_{0,TF}$, shows an even stronger dependence on the contact force, increasing by a factor of 4800, from 3.5 pA to 17 nA.

We find that both the ideality factor and the saturation current of the flake-substrate barrier remained unaltered by the load force, with values $I_{0,fs}$ = 52 nA and $n_{fs}$ = 1.01. The tip-flake barrier also exhibits a force-independent ideality factor, $n_{tf}$ = 1.9 but, on the other hand, its saturation current shows a dramatic dependence on the applied load, increasing by a factor of 4800, from 3.5 pA at 0 nN to 17 nA at 80 nN. Although we expect an increase of the tip-flake contact area with the applied load, which according to Equation 2 would cause a proportional increase of the tip-flake saturation current, this effect alone is insufficient to explain the huge change observed. Indeed, using a Hertzian contact model we estimated the change of the contact area due to the applied load (see Supp. Info.), finding that the contact area increases only by a factor of 4, much lower than the increase of a factor of 4800 observed in the saturation current. We believe that the dominant effect causing the increase of the saturation current is a lowering of the Schottky barrier formed at the tip-MoS$_2$ interface, probably due to a strain-induced reduction of the MoS$_2$ electron affinity [16, 36, 37]. Note that, according to Equation 2, $I_{0,\text{tf}}$ depends exponentially on the barrier height $\varphi_{\text{tf}}$ and, therefore, a small decrease of $\varphi_{\text{tf}}$ can cause a much higher increase of the saturation current. Replacing in Equation 2 the values of the saturation current extracted from the fit, we estimate the values of $\varphi_{\text{tf}}$ under the different applied loads, finding that the tip-flake barrier height changes by less than a factor 1.5 between 10 and 80 nN, from 313.4 meV at 10 nN to 213.4 meV at 80 nN.

Next, we measured *I-V* characteristics in a MoS$_2$ single layer (Figure 4a) and in a bilayer (Figure 4b) under different load forces. As in the case of the 8 layers thick flake, the *I-V*'s show a rectifying behavior, that can be reproduced by the theoretical model, although for such low semiconductor thicknesses the dominant transport mechanism could differ from the thermionic emission mechanism considered in the model [38, 39].

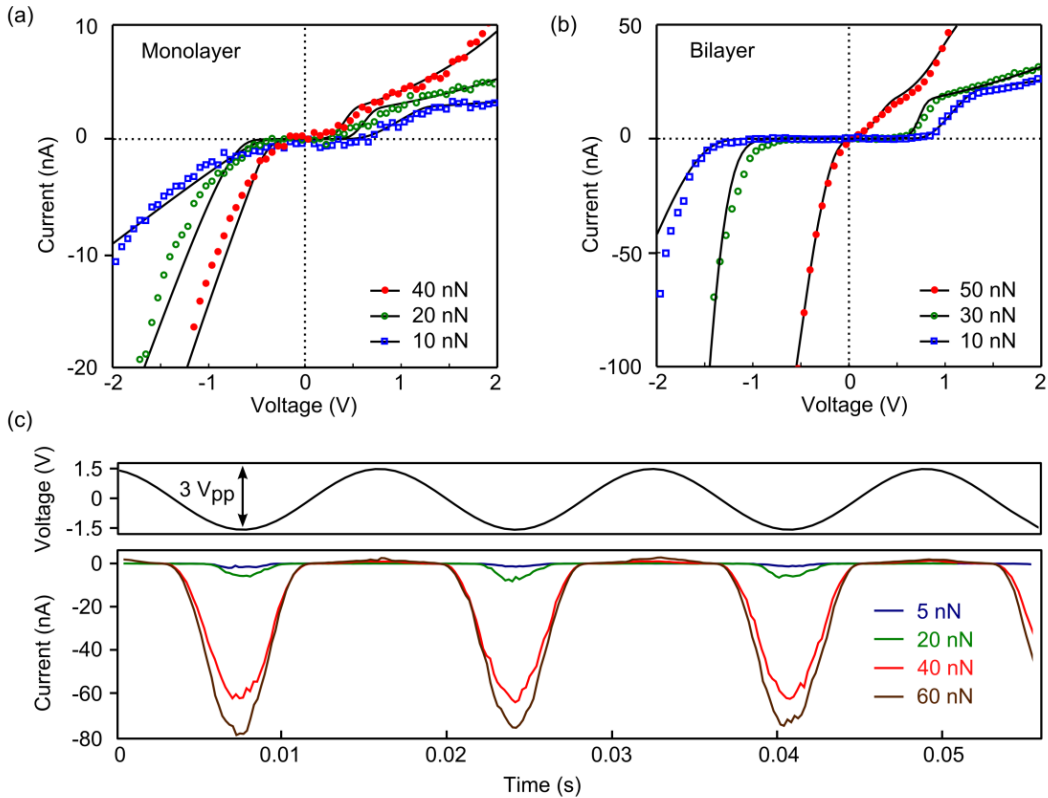

**Figure 4.** *I-V* characteristics of the MoS2 monolayer (a) and bilayer (b) vertical structures under three different tip-flake contact forces. Black lines are fits to the model. The flake-substrate saturation current, $I_{0,TF}$, decreases with the thickness of the MoS2 flake, from 52 nA in the 8-layer (see figure 3) to 18 nA in the bilayer and 2.8 nA in the monolayer, indicating an increase of the flake-substrate Schottky barrier height in the thinnest flakes. (c) Time-dependent current flow when a sinusoidal voltage (1.5 V, 60 Hz) is applied to the tip-monolayer MoS2-ITO structure. Depending on the load force the rectification ratio increases from 1 at 0 nN to 80 at 60 nN.

We find that the flake-substrate saturation current, $I_{0,fs}$, extracted from the model, decreases with the thickness of the MoS2 flake. At a load force of 10 nN we get $I_{0,fs} = 52$ nA in the multilayer, 18 nA in the bilayer and 2.8 nA in the monolayer. This reduction of $I_{0,fs}$ points to an increase of the Schottky barrier heights in the thinnest flakes, in concordance with the thickness dependence of the electron affinity of MoS2 described in recent literature [7, 39, 40]. The resistance of the flake, $R$, is also smaller in monolayers and bilayers. At 10 nN we get $R = 120$ kΩ for the monolayer and 11 kΩ for the bilayer, much lower than the 6500 kΩ obtained for the multilayer flake.

We measured *I-V*s at load forces ranging from 0 to 40 nN in the monolayer and from 0 to 50 nN in the bilayer. At higher load forces the flakes were usually damaged. We find that the series resistance, $R$, of the monolayer and bilayer flakes decreases monotonically when the load force is increased, as also observed in thicker flakes, ranging from 120 kΩ at 10 nN to 40 kΩ at 40 nN in the monolayer and from 11 kΩ at 10 nN to 1.9 kΩ at 50 nN in the bilayer. A similar piezoresistive behavior has been reported also for MoS$_2$ crystals subject to in-plane strain [16]. Further, in the case of MoS$_2$ crystals with an odd number of layers, in-plane strain also causes a piezoelectric response [41, 42] due to the broken inversion symmetry of 2H-MoS$_2$ layers [43]. Although a piezoelectric response could also emerge for MoS$_2$ crystals under out-of-plane strain, the I-V measurements presented in this work are not suitable to detect this effect, since they are carried out under stationary strain.

As for the 8-layer flake, in the case of monolayers and bilayers the ideality factors of the two barriers, $n_{fs}$ and $n_{tf}$, are independent of the load force. For the monolayer (bilayer) we get $n_{fs}$ = 1.5 (1.7) and $n_{tf}$ = 1.01 (1.02). In the cases of the monolayer and bilayer flakes we find again tip-flake saturation currents, $I_{0,tf}$, strongly dependent on the load force, however, in this case the flake-substrate saturation current, $I_{0,fs}$, is no longer force-independent, decreasing from 2.42 nA at 10 nN to 1.8 nA at 40 nN. This indicates that, for monolayers and bilayers, the local strain caused by the tip reaches the MoS$_2$-ITO interface, due to the lower thickness of the flake.

Finally, we demonstrate a practical example of application where the force-dependent electrical response of the studied structure is used to rectify a periodic signal. As shown in Figure 4, we were able to widely modulate the reverse current by changing the applied force, from ~1 nA to 75 nA while keeping the forward current stably rectified. The rectification ratio was therefore modified from a value of ~1 at a 5 nN load force to one as high as 80 at 60 nN.

It must be noted that the relation between the applied force and the rectification ratio of the device is not necessarily linear. In fact, as shown in figure 4c the forward current changes very rapidly at forces between 20 and 40 nN, and more slowly between 40 and 60 nN. A similar behavior is also observed in the 8-layer MoS$_2$ flake (see inset in Figure 3a), where the rectification ratio is strongly force-dependent between 20 and 40 nN, and remains almost constant in the range between 40 and 50 nN.

## 3. Conclusions

In summary, we studied the effect of indentation induced strain in the electron transport through vertical metal/atomically thin MoS$_2$/metal junctions for MoS$_2$ flakes as thin as a single layer. We measured the *I-V* characteristics of a vertical metal/MoS$_2$/metal structure with a conductive AFM tip and a metallic ITO substrates acting as electrodes. The resulting *I-V*s show a strong rectifying behavior, with a rectification ratio that can be modified when a load force is applied. The experimental *I-V*s were accurately reproduced by a double Schottky barrier model, and the saturation currents and ideality factors of the two barriers, as well as the MoS$_2$ resistance were extracted from the fit to the model. We find that the resistance of the MoS$_2$ flake decreases when a load force is applied, while the saturation current of the tip-MoS$_2$ barrier increases dramatically, indicating a pressure-induced lowering of the barrier potential. Finally, we used the metal/MoS$_2$/metal structure to rectify a sinusoidal voltage, controlling the rectification ratio with the applied load force. This work shows a very promising method to tailor the electronic properties of monolayer and few-layer MoS$_2$ vertical structures, which could be applied in the development of new atomically thin devices such as two-dimensional pressure sensors or mechanically tunable current rectifiers.

**Acknowledgments**

The authors acknowledge Victor Rubio for kindly providing indium tin oxide samples. G. R-B, N.A and J.Q acknowledge financial support from MICINN/MINECO through program MAT2014-57915-R, from Comunidad Autónoma de Madrid through program S2013/MIT-3007 (MAD2D) and from the European Commission under the Graphene Flagship, contract CNECTICT-604391. A.C-G. acknowledges financial support from the BBVA Foundation through the fellowship "I Convocatoria de Ayudas Fundacion BBVA a Investigadores, Innovadores y Creadores Culturales", from the MINECO (Ramón y Cajal 2014 program, RYC-2014-01406) and from the MICINN (MAT2014-58399-JIN). JJP acknowledges Spain's Ministerio de Economía y Competitividad for financial support under grant FIS2013-47328-C02-1 and the Generalitat Valenciana under grant no. PROMETEO/2012/011.

*Note added.*—During the preparation of this manuscript an article appeared showing mechanically tunable electrical performance in a vertical metal/semiconductor/metal atomically thin structure [44], where few-layer black phosphorus is used as semiconducting layer instead of the atomically thin MoS$_2$ layer considered in our work. Interestingly, while the *I-V* characteristics presented in their article almost remain unchanged when the sign of bias voltage is reversed, here we obtain markedly asymmetric *I-V* characteristics, leading to a strong rectifying behavior.

...

# Supporting information to: Strain engineering of Schottky barriers in single- and few-layer MoS$_2$ vertical devices.

Jorge Quereda[1], Andrés Castellanos-Gomez[2], Nicolás Agräit[1,2,3,4] and Gabino Rubio-Bollinger[1,3,4]

[1]Departamento de Fisica de la Materia Condensada. Universidad Autónoma de Madrid, Madrid, E-28049, Spain. [2]Instituto Madrileño de Estudios Avanzados en Nanociencia (IMDEA – Nanociencia), E-28049, Madrid, Spain. [3]Instituto de Ciencia de Materiales Nicolás Cabrera, E-28049, Madrid, Spain. [4]Condensed Matter Physics Center (IFIMAC), Universidad Autónoma de Madrid, E-28049, Madrid, Spain.

**Table of contents:**

1. Sample preparation.
2. Numerical solution for the back-to-back diode circuit.
3. Estimation of the contact area: Hertzian model.
4. Tip-ITO *I-V*s
5. Fitting parameters for the *I-V* characteristics.

**1. Sample preparation.**

We use a recently developed technique, based on the use of viscoelastic poly-dimethylsiloxane stamps [1], to transfer atomically thin MoS$_2$ crystals onto a transparent indium tin oxide (ITO) substrate. First, we use micromechanical cleavage to exfoliate MoS$_2$ crystals repeatedly until sufficiently thin flakes (1 to 15 atomic layers) are obtained. Then, we re-exfoliate these flakes using a poly-dimethylsiloxane viscoelastic stamp. We inspect the surface of the stamp using transmission optical microscopy and identify atomically thin MoS$_2$ flakes by their optical contrast ($C = 0.06 \pm 0.01$ in the case of monolayers) [2-4]. Next, the stamp and the ITO surface are gently put into contact and, subsequently, the stamp is slowly detached, leaving the few-layer MoS$_2$ flakes on top of the ITO substrate.

## 2. Numerical solution for the back-to-back diode circuit.

The Equation 5 in the main text cannot be solved analytically for $V_{\text{TF}}$, but requires numerical calculation instead. We recur to a method based on finding the intersection between $I_{\text{TF}}(V_{\text{TF}})$ and $I_{\text{FS}}(V_{\text{FS}})$, where $V_{\text{FS}} = V - V_{\text{TF}} - I_{\text{TF}}(V_{\text{TF}})R$. The Matlab™ code to solve the circuit is shown here:

```
function [J] = I(x0,V)
% I(x0,V) calculates the current flow through the back-to-back
% cirquit shown in figure 2c.
% x0 is a vector containing the parameters of the cirquit:
% x0(1) is the ideality factor n_TF
% x0(2) is the ideality factor n_FS
% x0(3) is the saturation current J_0,TF
% x0(4) is the saturation current J_0,FS
% x0(5) is the thermal voltage (K_b*T)/q (~0.025 at room temperature)
% x0(6) is the series resistance R

%% load parameters
if nargin == 0 %default parameters
    n1=1.5; %Ideality factor of I1
    n2=1.0; %Ideality factor of I2
    J01=1; %Saturation current for diode 1
    J02=100; %Saturation current for diode 2
    V_t=0.025; %(K_b*T)/q;
    R=1;
else %load
    n1=x0(1);n2=x0(2);J01=x0(3);J02=x0(4);V_t=x0(5);R=x0(6);
end

if nargin < 2
V=linspace(-1,1,100);
end

%% normalize saturation currents
J02copy=J02; % a copy of J02
J01=J01/J02; % normalize J01
R=R*J02; % Normalize R
J02=1; % Set J02 to 1

%% Calculate total current
for i=1:length(V)
    theV=V(i);
    options=optimset('TolX',1e-12,'Diagnostics','off', 'Display','off');
    output(:,i)=fsolve(@f,ones(3,1)*theV/3,options); % find the correct V1 (initial guess: Vi=V/3)
    J(i)=I1(output(1,i));
end
%% renormalize
```

```
J=J*J02copy;
R=R/J02copy;

%% nested functions
    function[F]=f(input) %F must be zero
        V1=input(1); %Voltage drop at D1
        V2=input(2); %Voltage drop at D2
        VR=input(3); %Voltage drop at R
        F(1)=I1(V1)-I2(V2);
        F(2)=R*I1(V1)-VR;
        F(3)=VR+V1+V2-theV;
    end
    function[I1]=I1(V1) % Shockley model for a Schottky diode (1)
        I1=J01*exp(V1./(n1*V_t)).*(1-exp(-V1./V_t));
    end
    function[I2]=I2(V2) % Shockley model for an inverted Schottky diode (2)
        I2=-J02*exp(-V2./(n2*V_t)).*(1-exp(V2./V_t));
    end
end
```

## 3. Estimation of the contact area: Hertzian model.

Using Hertzian contact mechanics we estimate the force-dependent contact area between the AFM tip and the MoS2 flake. We model the AFM tip as a sphere of radius R and the flake as a half-space, as shown in S1.

According to the Hertzian model the load force, $F$, and the indentation depth, $d$, are related by [5]:

$$F = \frac{4}{3} E^* R^{1/2} d^{3/2}, \qquad \text{(S-1)}$$

where $R$ is the curvature radius of the AFM tip (~100 nm for the HA-HR-DCP conductive diamond tips from NT-MDT) and $E^*$ is a function of the Young moduli ($E_T$ and $E_F$), and the Poisson ratios, ($\nu_T$ and $\nu_F$) of the diamond AFM tip and the MoS2 flake, respectively:

$$E^* = \left(\frac{1-\nu_T^2}{E_T} + \frac{1-\nu_F^2}{E_F}\right)^{-1}, \qquad \text{(S-2)}$$

with values [6, 7] $\nu_T = 0.2$, $E_T = 1200$ GPa, $\nu_F = 0.27$, and $E_F = 0.83$ GPa.

The radius of the contact area is related with the applied force by

$$a = \left(\frac{3FR}{4E^*}\right)^{1/3}, \qquad (S\text{-}3)$$

and the tip-flake contact area $A_c$ is

$$A_c = \pi(a^2 + d^2) = \pi\left\{\left(\frac{3FR}{4E^*}\right)^{2/3} + \frac{1}{R^2}\left(\frac{3FR}{4E^*}\right)^{4/3}\right\}. \qquad (S\text{-}4)$$

The distribution of normal pressure, $P$, as a function of distance from the center of the contact area, $r$, is given by

$$P = P_0\left(1 - \frac{r^2}{a^2}\right)^{\frac{1}{2}}, \qquad (S\text{-}5)$$

where

$$P_0 = \frac{1}{\pi}\left(\frac{6FE^{*2}}{R^2}\right)^{\frac{1}{3}}. \qquad (S\text{-}6)$$

Figure S1b shows the tip-flake contact area and indentation depth as a function of the applied force. For forces between 10 and 80 nN the contact area increases by roughly a factor of 4, from 281 nm² to 1157 nm². According to Equation 2 (main text), the saturation current, $I_{0,\mathrm{TF}}$, is proportional to the contact area and, therefore, it should also increase by a factor of 4 when the applied force is increased from 10 nN to 80 nN. However, the observed increase of $I_{0,\mathrm{TF}}$ is much higher (a factor of 4800), indicating that this change is dominantly caused by a force-induced lowering of the Schottky barrier formed at the tip-MoS$_2$ interface, and not by the increase of the contact area.

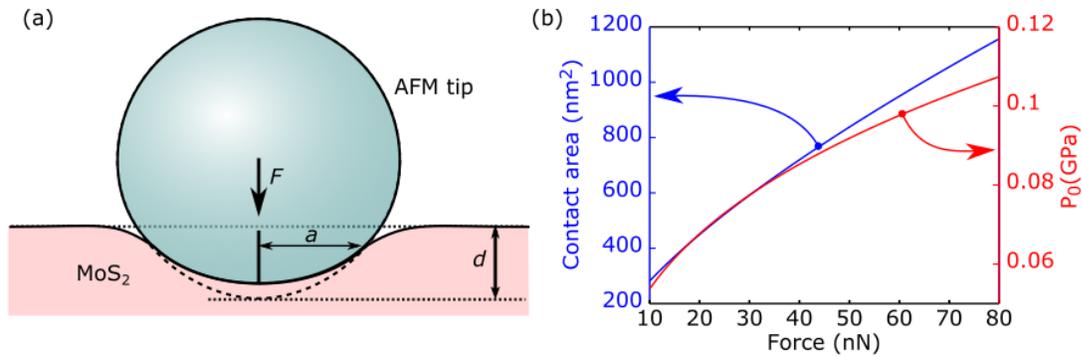

**Figure S1.** (a) Hertzian contact model of the tip-flake interface. The AFM tip is modeled as a sphere and the MoS₂ flake as a semi-space. $F$ is the applied load force, $a$ is the radius of the contact area and $d$ is the indentation depth. (b) Tip-flake contact area and pressure under the AFM tip as a function of the applied force.

## 5. Estimation of the force-dependent tip-MoS$_2$ Schottky barrier heights.

Using the contact areas calculated in the previous section, we estimate the force-dependent tip-flake Schottky barrier heights for the different MoS$_2$ thicknesses. Reordering the Equation 2 in the main text one gets

$$\varphi_i = \frac{k_B T}{q} \cdot \ln\left(\frac{I_0}{A_c A^* T^2}\right). \quad \text{(S-7)}$$

Where the Richardson effective constant, $A^*$, is given by

$$A^* = \frac{4\pi m^* q k_B^2}{h^3} = 1.20173 \times 10^6 \, Am^{-2} K^{-2} \times \frac{m^*}{m_e} \quad \text{(S-8)}$$

Replacing $I_0$ with the values extracted from the fitting of the *I-V* characteristics and considering $m^*/m_e \approx 0.5$ [8] one gets the estimated Schottky barrier heights.

Figure S2 shows the estimated tip/MoS$_2$ Schottky barrier height as a function of the indentation force for the 8-layer flake. As discussed in the main text, the tip-flake Schottky barrier decreases as the force is increased, from 313.4 meV at 10 nN to 213.4 meV at 80 nN.

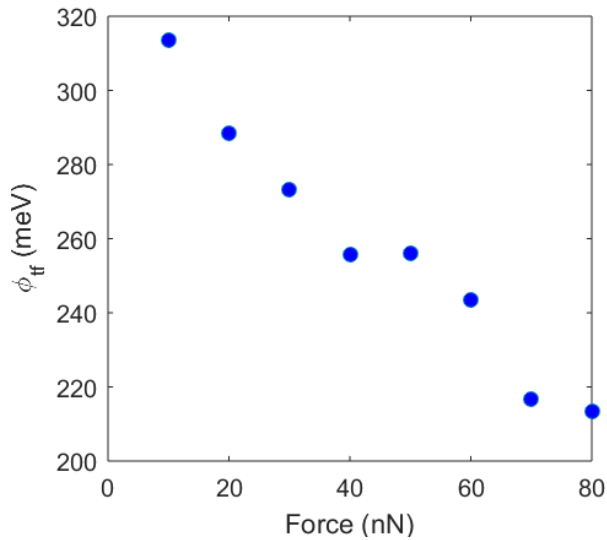

**Figure S2.** Estimated force-dependent tip/MoS$_2$ Schottky barrier heights for the 8 layers thick MoS$_2$ crystal.

The estimated thickness-dependent tip/MoS$_2$ Schottky barrier heights for monolayer and bilayer MoS2 are presented in the Table S1. Again, in both cases the barrier decreases monotonically as the force is increased.

Table S1. Estimated force-dependent tip-flake Schottky barriers for the MoS$_2$ monolayer and bilayer.

| Monolayer | | Bilayer | |
|---|---|---|---|
| Force (nN) | $\varphi_{tf}$ (meV) | Force (nN) | $\varphi_{tf}$ (meV) |
| 10 | 1663 | 10 | 1809 |
| 20 | 1620 | 30 | 1715 |
| 40 | 1500 | 50 | 1365 |

## 6. Tip-ITO *I-V*s

Figure S3 shows an *I-V* characteristic measured with the conductive AFM tip directly on top of the ITO substrate (blue line). The resulting *I-V* is ohmic, in contrast with the rectifying *I-V* obtained in the tip-MoS$_2$-ITO structure (red line). We conclude that the parasitic resistances caused by the diamond tip and the ITO substrate have a negligible effect on the *I-V* curve of the tip-MoS$_2$-ITO structure compared with the dominant effect of the MoS$_2$ resistance and the tip-MoS$_2$ and MoS$_2$-ITO interfaces.

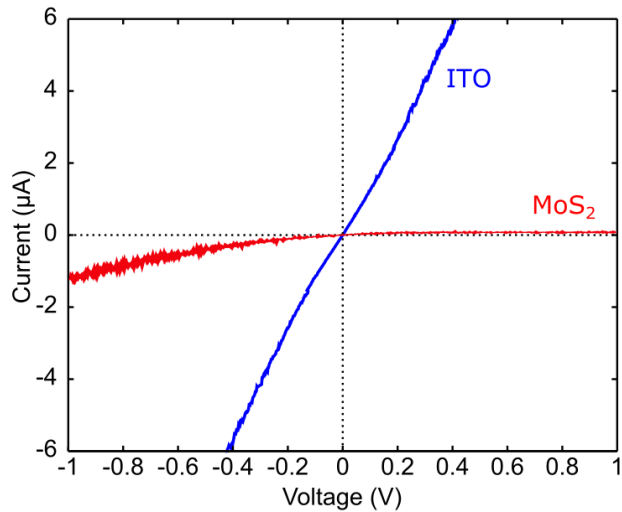

**Figure S3.** Comparison between I-V characteristics measured placing the AFM tip on top of the ITO substrate (blue) and on top of a 6 nm thick MoS$_2$ crystal (red). The same load force (80 nN) is applied in both cases.

## 7. Transmission line model of current injection into the MoS$_2$ crystal.

We further characterize the distribution of the current inside the MoS$_2$ crystals using a transmission line model [9, 10], where the crystals are represented as three-dimensional resistor network, as the one shown in Figure S3a. According to literature, the in-plane ($\sigma_\parallel$) and out-of plane ($\sigma_\perp$) conductivities of MoS$_2$ are related by $\sigma_\parallel = 200\,\sigma_\perp$, where, at room temperature, $\sigma_\parallel = 0.03\,\Omega^{-1}\,\text{cm}^{-1}$ [11, 12]. Therefore, we consider different values for the in-plane ($R_\parallel$) and out-of-plane ($R_\perp$) resistances, where $R_\perp = 200\,R_\parallel$. The electronic contacts made by the AFM tip and the ITO substrate are included in the model by fixing the voltage at the central node of the top layer of the network to $V_{\text{tip}}$, while keeping nodes of the bottom layer grounded, as shown in Figure S3-a. In order to characterize the spreading of the current inside the MoS$_2$ crystal we calculated the current at the different layers of out-of-plane resistances, which we indexed from *k*=1 (top layer) to *k*=10 (bottom layer). We find that, as the layer index *k* increases, the current is spread along progressively wider areas below the position of the AFM tip. Figure S3b shows the flow of current at a circular region below the AFM tip as a function of the area of the considered region. In the first layer ($k=1$, red curve) the flow of current is markedly concentrated below the AFM tip, and increases rapidly with the area in the vicinity of the tip. However, for the subsequent layers the current becomes progressively more spread in the crystal. The inset in Figure S3b shows the area $A_{75\%}$ at which the out-of-plane current reaches the 75% of $I_0$, in each layer, as a function of the vertical position, $k$. For the first layer below the tip we get $A_{75\%} = 0.1 A_T$, where $A_T$ is the total area of each layer, $A_T = 200 \times 200$. For the last layer, however, the current is distributed in a wider region and we get $A_{75\%} \approx 0.7 A_T$. It must be noted that, according to the model, even in the monolayer crystal the current spreads horizontally along a remarkably wide area. In fact, even though the bias voltage is only applied to the central node of the top network layer, we get $A_{75\%}(\text{monolayer}) = 4000$. Therefore, it is justified to say that even for a monolayer the Schottky barriers are not formed on top of each other.

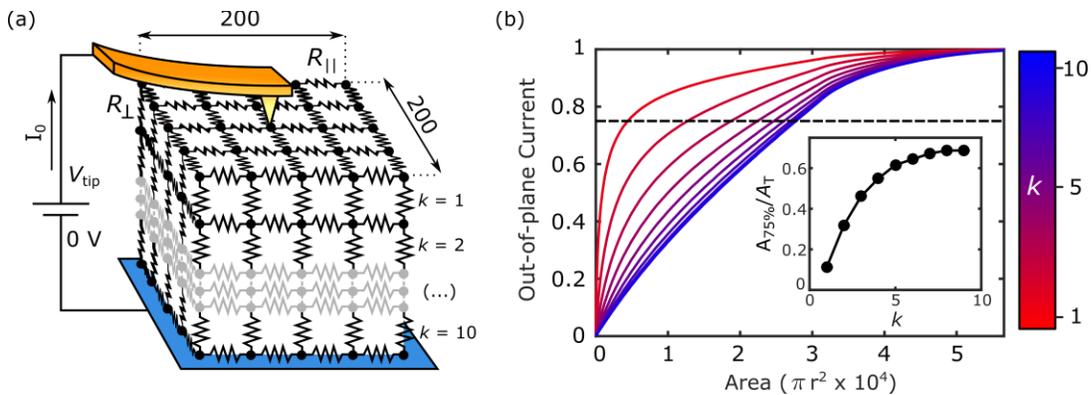

**Figure S4.** (a) The MoS$_2$ crystal is modeled as a three-dimensional resistor network with in-plane resistances $R_\parallel$ and out-of-plane resistances $R_\perp$. The bottom surface of the cube is grounded while the central node of the top surface is fixed to a voltage $V_{\text{tip}}$. The calculations are performed considering a network of 200×200×10 resistances with $V_{\text{tip}} = 10$ V, and $R_\perp = 200\,R_\parallel$. (b) Current flow in the out-of-plane direction through a circular region centered at the horizontal position of the tip, as a function of the radius of the considered region. Each curve corresponds to a different horizontal layer in the resistor network, from top (red) to bottom (blue). The values of the current are normalized to the total out of plane current, $I_0$. Inset: Area $A_{75\%}$ at which the current equals the 75% of $I_0$, as a function of the layer depth.

## 8. Stability of the force-dependent *I-V* curves

Figure S5 shows two sets of force-dependent *I-V* curves acquired consecutively in the same region of the 8-Layer MoS$_2$ flake to prove the stability of the tip/MoS$_2$/ITO structure.

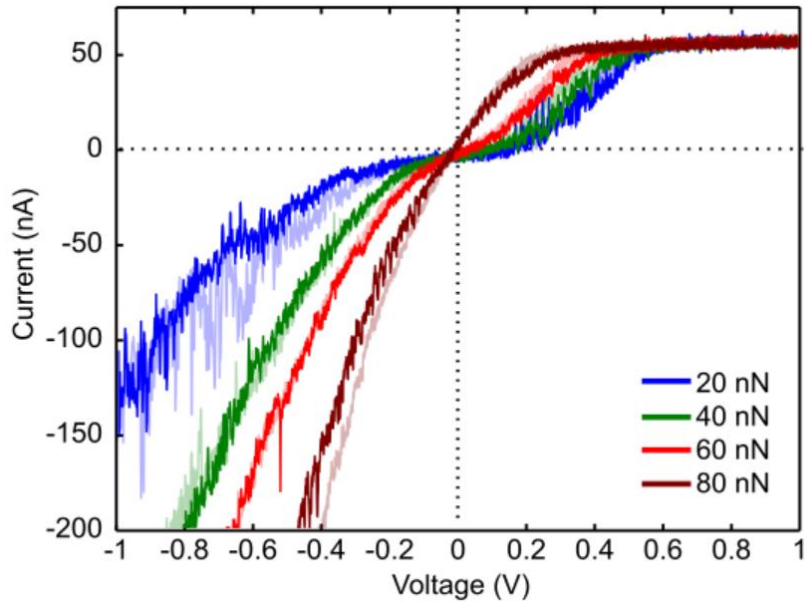

**Figure S5.** Force-dependent *I-V* characteristics of the 8 layers thick MoS$_2$ flake under four different tip-flake contact forces: 20, 40, 60 and 80 nN. A second set of measurements, acquired consecutively, is shown in smooth colors to check the stability of the *I-V* characteristics.

## 9. Fitting parameters for the *I-V* characteristics.

Monolayer:

| $n_{TF}$ | $n_{FS}$ | $I_{0.TF}$ (A) | $I_{0.FS}$ (A) | $V_t$ (V) | $R$ (Ω) | $F$ (nN) |
|---|---|---|---|---|---|---|
| 1.93 | 1.01 | 1.71e-15 | 2.42e-9 | 0.026 | 1.19e8 | 10 |
| 1.58 | 1.01 | 1.44e-14 | 1.78e-9 | 0.026 | 5.95e7 | 20 |
| 1.99 | 1.04 | 2.45e-12 | 1.80e-9 | 0.026 | 4.08e7 | 40 |

Bilayer:

| $n_{TF}$ | $n_{FS}$ | $I_{0.TF}$ (A) | $I_{0.FS}$ (A) | $V_t$ (V) | $R$ (Ω) | $F$ (nN) |
|---|---|---|---|---|---|---|
| 1.65 | 1.01 | 6.09e-18 | 1.92e-8 | 0.026 | 1.13e7 | 10 |
| 1.76 | 1.01 | 3.60e-16 | 1.82e-8 | 0.026 | 2.65e6 | 30 |
| 2.13 | 1.05 | 4.48e-10 | 1.70e-8 | 0.026 | 1.94e6 | 50 |

8-Layer:

| $n_{TF}$ | $n_{FS}$ | $I_{0.TF}$ (A) | $I_{0.FS}$ (A) | $V_t$ (V) | $R$ (Ω) | $F$ (nN) |
|---|---|---|---|---|---|---|
| 1.61 | 1.00 | 1.62e-08 | 4.80e-08 | 0.027 | 7.01e5 | 80 |
| 1.54 | 1.01 | 1.31e-08 | 4.52e-08 | 0.027 | 7.47e5 | 70 |
| 1.64 | 1.01 | 4.15e-09 | 4.79e-08 | 0.027 | 9.32e5 | 60 |
| 2.32 | 1.00 | 2.26e-09 | 5.27e-08 | 0.027 | 1.42e6 | 50 |
| 1.95 | 1.00 | 1.97e-09 | 5.30e-08 | 0.026 | 1.39e6 | 40 |
| 2.25 | 1.00 | 8.23e-10 | 5.31e-08 | 0.025 | 2.60e6 | 30 |
| 1.86 | 1.00 | 3.48e-10 | 5.54e-08 | 0.027 | 4.59e6 | 20 |
| 1.73 | 1.00 | 8.26e-11 | 5.57e-08 | 0.027 | 6.18e6 | 10 |
| 2.40 | 1.00 | 3.50e-12 | 5.30e-08 | 0.027 | 8.00e6 | 0 |